\documentclass[a4paper,11pt]{article}
\usepackage{pos}

\usepackage{bibspacing}
\title{All sky archival search for FRB high energy counterparts with Swift and Fermi}
 \ShortTitle{FRB-GRB counterparts}

\author*[1]{H. Ashkar}
\author[1]{M. El Bouhaddouti}
\author[1]{S. Fegan}
\author[2]{F. Sch\"ussler}
\affiliation[1]{Laboratoire Leprince-Ringuet, École Polytechnique, CNRS, Institut Polytechnique de Paris, F-91128 Palaiseau, France}

\affiliation[2]{IRFU, CEA, Universit\'e Paris-Saclay, F-91191 Gif-sur-Yvette, France}

\emailAdd{halim.ashkar@llr.in2p3.fr}

\abstract{Fast radio bursts (FRBs) are millisecond-duration radio signals from unknown cosmic origin. Many models associate FRBs with high-energy astrophysical objects such as magnetars. In this attempt to find counterparts to FRBs, we explore gamma-ray bursts (GRBs) from the Swift and Fermi missions. We first search for spatial correlations between FRB and GRB populations as a whole and then search for a one-by-one correlation between each of the FRBs and GRBs investigated. Temporal coincidences are not considered. To evaluate the significance of any correlation found, we generate background realizations that take into account instrumentally induced anisotropies in the distribution of the sources. Neither study yields any significant counterpart detection. We estimate that less than 4\% of the FRBs are associated with GRBs in the studied samples.}

\ConferenceLogo{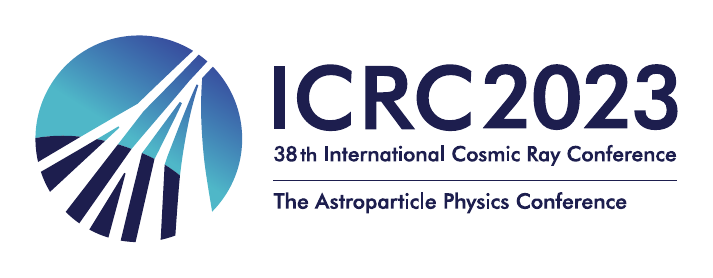}

\FullConference{%
38th International Cosmic Ray Conference (ICRC2023)\\
  26 July - 3 August, 2023\\
  Nagoya, Japan}

\begin{document}
\maketitle

\section{Introduction}
\label{sec:introduction}
Fast Radio Bursts (FRBs) are flashes of radio emission from astrophysical origin that last from a fraction of a millisecond to a few milliseconds. The Lorimer Burst~\citep{Lorimer_2007} was the first FRB discovered using the archival data of the Parkes radio telescope. Since then, FRB detections have increased and reached more than 600 detections by July 2022. There are two types of FRBs: repeaters and non-repeaters. In 2020, for the first time, an FRB from galactic origin, FRB\,20200428A, was detected~\citep{SGR1935_CHIME, SGR1935_CHIME2, SGR1935_STAR2} repeatedly~\citep{SGR1935_FAST, Kirsten_2020}. The FRB is associated to a soft gamma-ray repeater (SGR), SGR1925+2154~\citep{Zhou:2020} which is a magnetar. FRB\,20200428A was detected during an active outburst phase of the magnetar that lasted several weeks. Coincidentally with the FRB, X-ray outbursts have been detected by several instruments~\citep{tavani2020xray, Ridnaia:2021,li2020insighthxmt, INTEGRAL_BURST_A}. A hint of a gamma-ray transient was discovered in the Swift satellite data contemporaneously to FRB\,20131104~\citep{DeLaunay_2016}. However, the 2022 outburst remains the only confirmed electromagnetic counterpart to an FRB until now (October 2022).

The short emission time and the high-temperature brightness of FRBs imply small emission regions and coherent processes~\citep{Petroff:2019tty}. Some of the most prominent sources linked to FRBs, and repeating FRBs in particular, are magnetars~\citep{Popov:2007uv,2017ApJ...843L..26B}. For example, the interaction of a magnetar with a surrounding nebula could produce FRBs through synchrotron maser emission~\citep{Lyubarsky:2014jta, Metzger:2019, Beloborodov:2020}. Some of these models suggest the FRB could be accompanied by a gamma-ray outburst. FRB\,20200428A, in spite of having a low energy budget on the FRB scale, is the first observational evidence of this association. For this specific FRB, it has been suggested that the unusually hard spectrum implies a common origin for the radio and X-ray emission~\citep{Ridnaia:2021}. The hard X-ray spectrum points to a non-thermal nature, which can lead to the production of gamma rays~\citep{li2020insighthxmt}. Other prominent sources linked to FRBs are neutron star interactions, hyperflares and giant flares from magnetars~\citep{PP13, PPP+18}, binary white dwarf mergers \citep{KIM13}, black hole interactions~\citep{LRL+16, Falcke_2014} and neutron star interactions~\citep{Totani13, Lyutikov_2013, YTK+18, VRB+17, Zhang18}. Whether the sources of FRBs are magnetars or compact object interactions, they are undoubtedly very energetic sources that would be capable of producing gamma rays. Magnetars for example are linked to short gamma-ray bursts ~\citep[GRBs,][]{burns2021identification} as well as mergers involving at least one neutron star~\citep{1989Natur.340..126E, 2015ApJ...815..102F}.

Several attempts have been made to search for FRB counterparts either by looking in archival data or by actively observing or following-up FRBs with no clear success until now. From recent archival searches no clear relations were found between IceCube neutrinos and CHIME FRBs~\citep{https://doi.org/10.48550/arxiv.2112.13820} and no association between gravitational waves and FRBs was found~\citep{https://doi.org/10.48550/arxiv.2203.17222, https://doi.org/10.48550/arxiv.2203.12038}. An optical counterpart was possibly found for FRB\,20180916b~\citep{Li_2022}. On the gamma-ray transients side, a possible counterpart for FRB\,20171209 was found~\citep{Wang_2020} from a search including 110 FRB and 1440 GRBs with an afterglow detection. Moreover, a gamma-ray transient has been reported as a possible counterpart to FRB\,20121104~\citep{2016ApJ...832L...1D}. A search including CHIME FRBs and GRBs (short and long) detected between July 2018 and July 2019 did not find any GRB counterparts to FRBs~\citep{https://doi.org/10.48550/arxiv.2208.00803}. A similar result was found for Insight-HXMT gamma-ray transients and FRBs~\citep{2020A&A...637A..69G}.


Our work aims to independently establish a spatial link between FRBs and gamma-ray transients, notably GRBs, by looking at archival data of the last two decades from active GRB observatories such as Swift and Fermi. Several models expect different timescales for FRB emission from GRB progenitors. For example, in some models of FRBs from young magnetars, the time difference between the initial cataclysmic event (GRB progenitor) and the FRB emission can be years or even decades~\citep{Metzger_2017, Murase_2016}. On the other hand, in the case of FRB\,20200428A the high energy emission arrived simultaneously with the FRB. In a merger-driven explosion scenario, it is possible that the FRB emission precedes the high energy emission from a long GRB~\citep{Dong_2021}. For these reasons, we do not consider any temporal constraints on the emission in our study. Since both neutron star mergers and the core collapse of massive stars could lead to the production of a magnetar, we do not differentiate between short and long GRBs. Finally, given that a fraction of  non-repeater FRBs might be undetected repeaters~\citep{10.1093/mnras/stac2174} and that the difference between repeaters and non-repeaters is not perfectly clear, we do not differentiate between these two types either.  


\section{Data and background}
\label{sec:Data}
For this study, we consider all FRBs since the first detection until July 2022. This sample of FRBs is taken from the Transient Name Server (TNS)\footnote{\url{https://www.wis-tns.org}}. For GRBs, we consider those detected by the Burst Alert Telescope (BAT) on board the  Swift obsevatory~\citep{2004ApJ...611.1005G} for their precise localisation~\citep{Lien_2016}. To avoid duplication, when the X-ray Telescope (XRT) position is available we only consider it. We also enlarge our sample by adding Fermi GRB detections. The Fermi GRBs are taken from the Gamma-ray Burst Monitor (GBM)~\citep{Meegan_2009} GRB catalog that is the catalog~\citep{von_Kienlin_2020, Gruber_2014, von_Kienlin_2014, Bhat_2016}. Moreover, we include GRBs from the Large Area Telescope~\citep[LAT,][]{Cecchi_2008,Abdollahi_2020} on the Fermi gamma-ray space telescope. We remove all entries from the Fermi catalogs that are also detected by Swift. Fermi-GBM GRBs are poorly localized with localization regions that span several square degrees in the sky. Therefore, we consider Fermi-GBM GRBs with localization uncertainties smaller than 1 degree separately. These are mainly detected and localized by other instruments. The three datasets represent GRBs detected in different energy ranges with different instrumental detection biases. They are first treated separately to highlight these differences, then combined into one dataset including all selected GRBs. 


The data is compared to a background generated from the data itself. While the underlying distribution of GRBs can be approximated as isotropic, the distribution of detected positions can be affected by instrumental biasses. Therefore, to simulate the background we use the positions of the sources in the dataset. The goal is to generate 1000 simulated GRB catalogues with positions that follow the distribution of the Swift-BAT GRBs in the sky. The separations squared between FRBs and GRBs are calculated for each simulated catalogue, binned and plotted in histogram after averaging the bins. 
Equidistant elements from the Swift-BAT declinations cosines are taken to generate a function describing their distribution.  The function is then applied to a set of random numbers between 0 and 1 having the same length as the initial sample to generate the cosine of the simulated declinations. The right ascensions are generated by subdividing the declination cosine distribution into 20 intervals and applying the same method as above for each of the sub-intervals. 


\section{Searching for gamma-ray counterparts for FRB and GRB populations}
\label{sec:PopulationFRB}

\begin{figure*}[!th]
  \centering
  \begin{minipage}[b]{0.40\textwidth}
    \includegraphics[width=\textwidth]{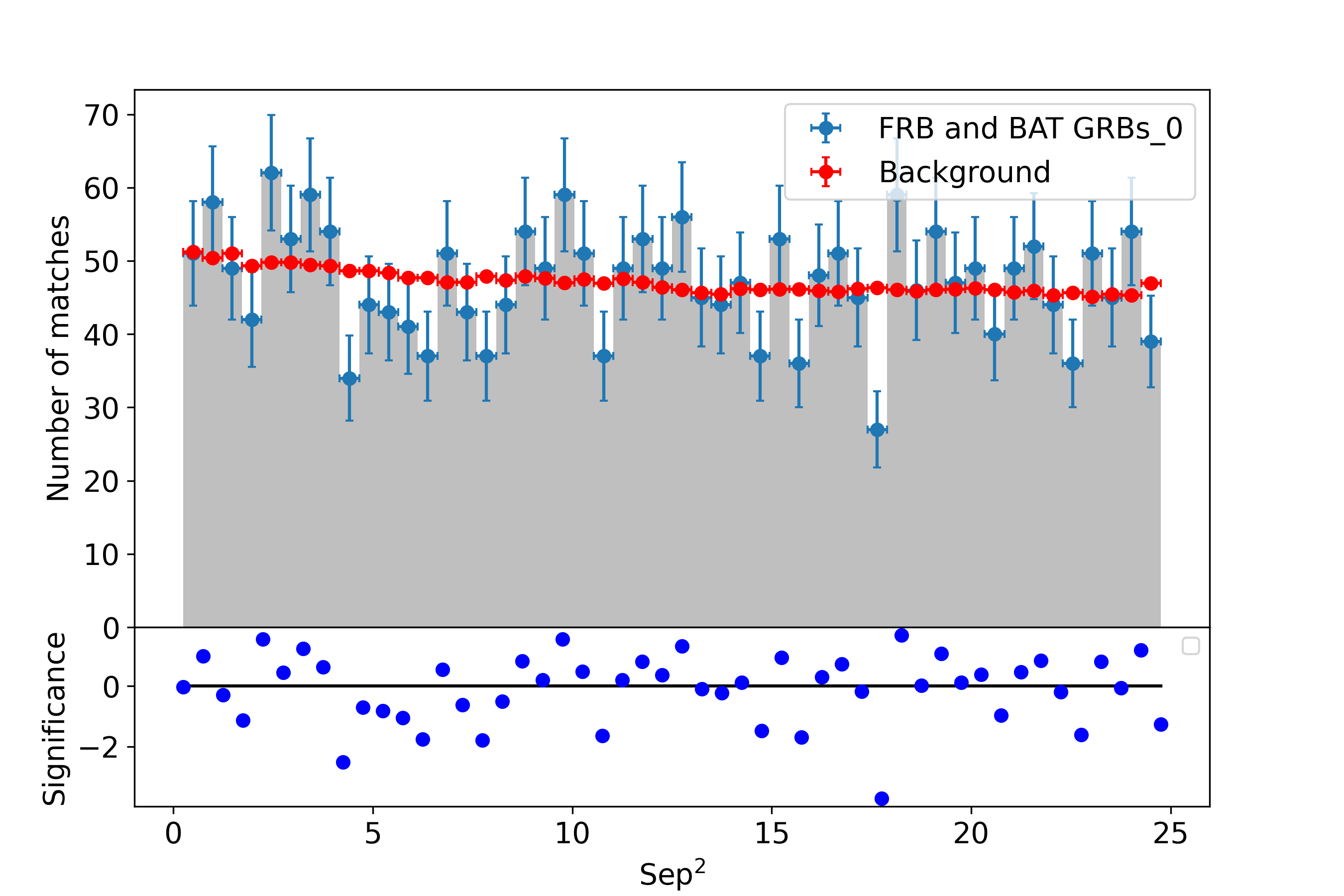}
  \end{minipage}
    \begin{minipage}[b]{0.40\textwidth}
    \includegraphics[width=\textwidth]{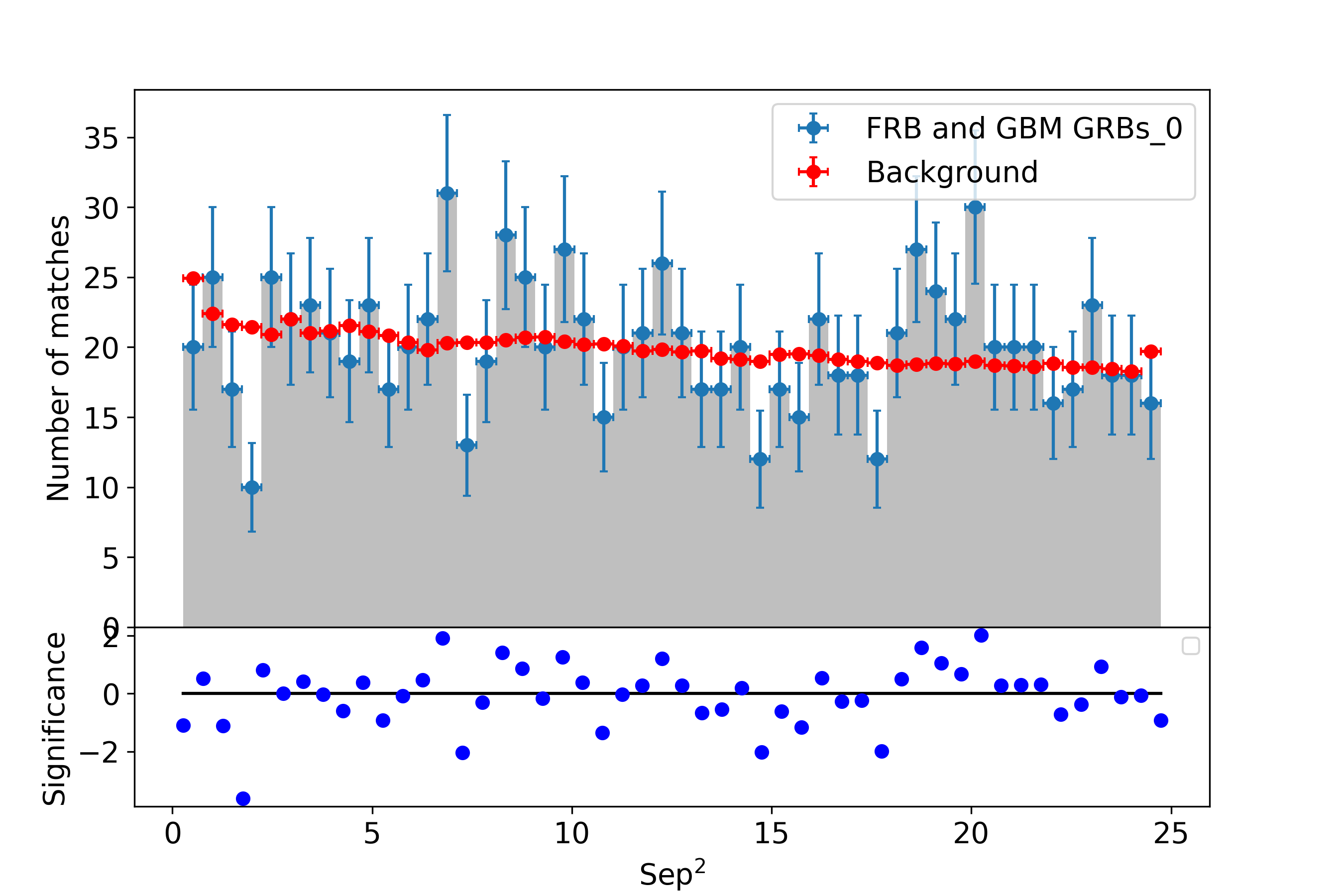}
  \end{minipage}
     \begin{minipage}[b]{0.40\textwidth}
    \includegraphics[width=\textwidth]{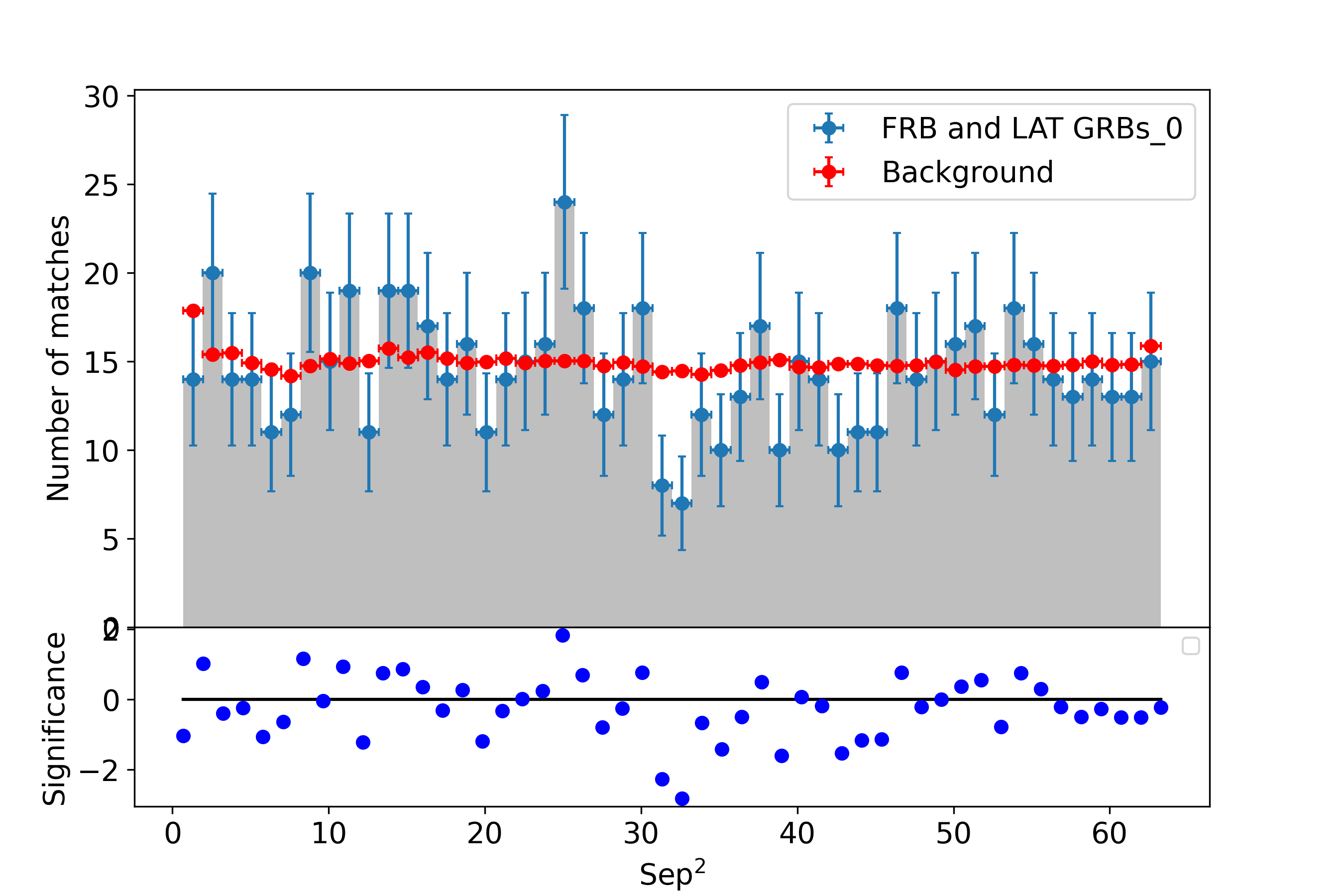}
    \end{minipage}
    \begin{minipage}[b]{0.40\textwidth}
    \includegraphics[width=\textwidth]{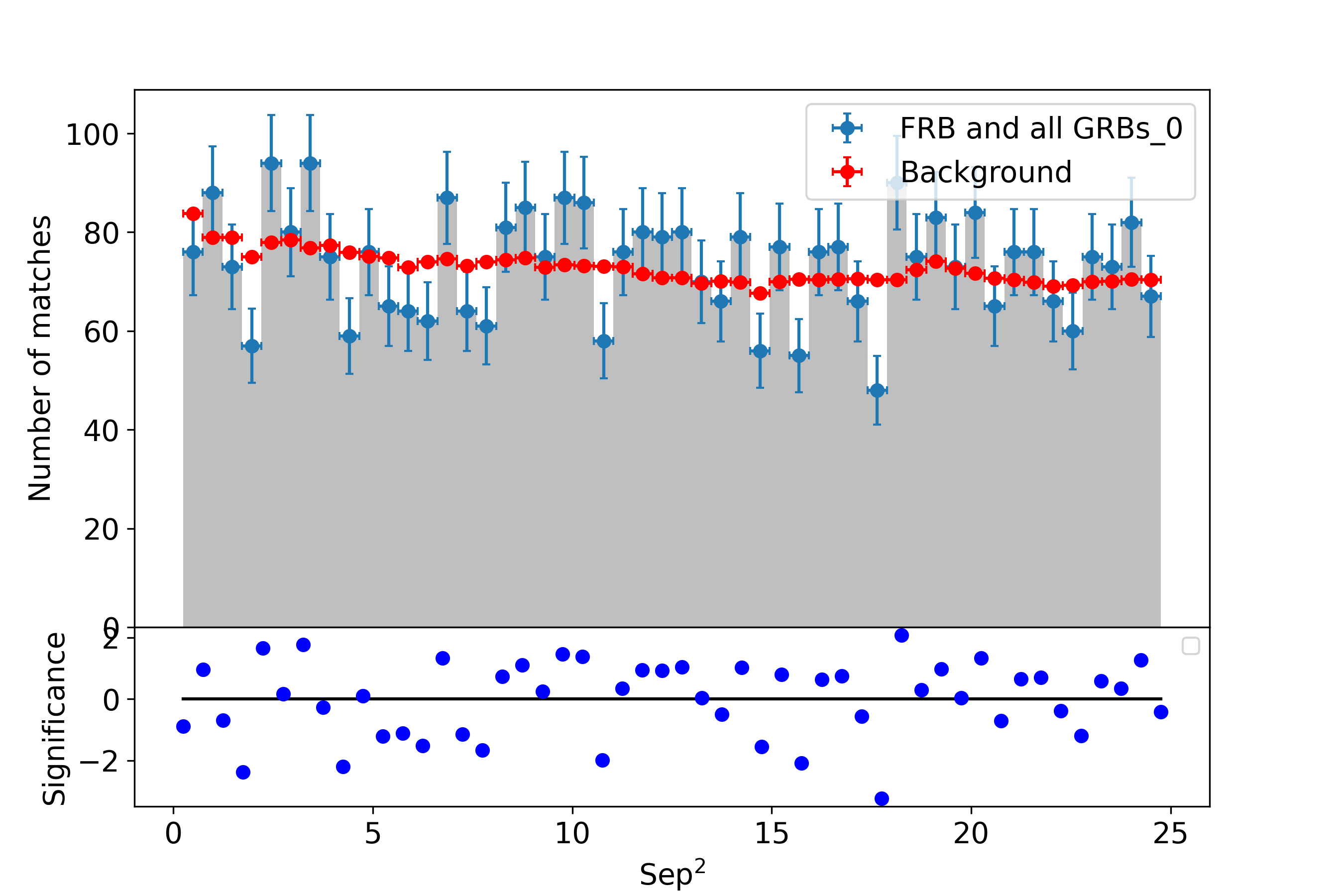}
      \end{minipage}
     \caption{$\mathrm{sep^2}$ between FRBs and Swift-BAT (upper left), Fermi-GBM ($<$ 1 deg uncertainty, upper right), Fermi-LAT (lower left), and the combination of all three GRB catalogs (lower right) distribution. The blue dots show the number of matches in each bin and their uncertainty (signal) and the red dots show the same for the averaged generated background (background). The bottom plots show the significance of the excess computed for each bin.}
     \label{fig:hist_frb_grb}
\end{figure*}

Populations of FRBs and gamma-ray transients can be studied to test spatial correlations between them~\citep{https://doi.org/10.48550/arxiv.2112.13820}. The idea here is to calculate separations squared ($\mathrm{sep^2}$) between FRBs and other transients, plot the histograms of separations squared and compare them to a background. If there is any spatial correlation between FRBs and other transients populations, we would expect to see a significant excess in the first few histogram bins. 

The background is generated as explained in Sec.~\ref{sec:Data}. The separation squared between FRBs and  GRBs is calculated. To determine the bin size, the average positional uncertainty is considered for FRBs and GRBs. The average error squared is: $\mathrm{\langle \delta^2 \rangle=\langle \delta_{FRB}^2 \rangle + \langle \delta_{GRB}^2 \rangle}$

We generate 1000 realizations. The separation squared are binned and the average of the bins are calculated. The resulting histograms are shown in Fig.~\ref{fig:hist_frb_grb}. The excess and the significance of the excess in each bin is: 
\begin{equation}
\mathrm{Excess=N_{On} - \alpha N_{Off}}
\quad\text{and}\quad
\mathrm{\sigma=\frac{N_{On}-\alpha N_{Off}}{\sqrt{N_{On}+\alpha^2 N_{Off}}}}
\end{equation}
where $\mathrm{N_{On}}$ is the number of entries for each signal bin, $\mathrm{N_{Off}}$ is the sum of entries of the background bins and $\mathrm{\alpha}$ is 1/1000.

This method is used to establish correlations between FRBs and Swift-BAT, Fermi-GBM ($<$ 1 deg uncertainties) and Fermi-LAT GRBs. The study is also repeated combining all three GRB catalogues. The bin size depends on the localization uncertainties. We find that for the comparison with Swift-BAT and Fermi-GBM, a bin size of 0.5 deg$\mathrm{^2}$ is suitable, allowing any excess to appear in the first bins if any significant correlation exists. Due to a small number of GRBs detected by Fermi-LAT, the bin size is slightly larger in order to increase the statistics in each bin. The simulated background follows the signal distribution which is a good indication of the reliability of the background generation method.  No significant excess can be extracted from the first few bins.


\section{Searching for gamma-ray counterparts for FRBs case by case}
\label{sec:CaseByCaseFRB}

\begin{table*}[!ht]
\centering
\small
\begin{tabular}{ccccccc}
  \hline
    Catalogues & FRB & GRB & Separation & $P_1$ & P \\  
        \hline
     FRB - BAT & FRB\,20190304C & GRB\,201128A &  0.057 & 0.028   & 1 \\
     FRB - BAT & FRB\,20171209A & GRB\,110715A &  0.084 & 0.005   & 1 \\
     FRB - GBM & FRB\,20181218A & GRB\,20306761 &  0.152 & 0.1734   & 1 \\
     FRB - GBM & FRB\,20181018C & GRB\,210308276 &  0.429 & 0.0132  & 1 \\
     FRB - LAT & FRB\,20190612A & GRB\,151006413 &  0.237 & 0.041  & 1 \\
     FRB - LAT & FRB\,20190201A & GRB\,211023546 &  0.805 & 0.033  & 1 \\
     FRB - ALL & FRB\,20190304C & GRB\,201128A &  0.057 &  0.044  & 1 \\
     FRB - ALL & FRB\,20171209A & GRB\,110715A &  0.084 & 0.007  & 1 \\
     
   \hline
   \hline
\end{tabular}
\caption{Smallest separations and chance probabilities obtained for FRB and GRB pairs. }
\label{tab:FRB-GRB-SEP}
\end{table*}


As a second attempt to establish a link between FRBs and GRBs we look at spatial coincidences between all well-localized GRBs and FRBs. 
Following~\cite{Wang_2020} and~\cite{Li_2022}, for each FRB, a 10-degree radius region around it is considered. The effective expected number of GRBs inside each of these areas is: $\mathrm{\lambda = \rho_i S}$, where $\mathrm{\rho_i}$ is the effective density of GRBs in the region around each of the FRBs and $S$ is the surface of the region. Taking into consideration the whole sky: $\mathrm{S = 41252.96(1-cos(D+\delta_{FRB}+\delta_{GRB}))~deg^{2}}$ where $\mathrm{D}$ is the angle between the FRB and the GRB, $\mathrm{\delta_{FRB} = \sqrt{\delta_{RA,FRB}\delta_{Dec,FRB}}}$ is the error radius of the FRB and $\mathrm{\delta_{GRB}}$ is the error radius of the GRB.
The GRBs inside the 10-degree radius region follow a Poisson distribution. The chance probability of finding one or more GRBs in the FRB test region is $\mathrm{P_{1,i}}$ and the post-trial chance probability $\mathrm{P}$ are:
\begin{equation}
\mathrm{P_{1,i} = 1 - \mathrm{exp(-\lambda)}}
\quad\text{and}\quad
\mathrm{P =1 - \prod_{i=1}^N (1- P_{1,i})}
\end{equation}
where $\mathrm{N = 627}$ is the total number of FRBs in our study. $\mathrm{P_{1,i}}$ and $\mathrm{P}$ values are computed for all the matches between CHIME FRBs and GRBs. 

We search for coincidences between FRBs and Swift-BAT, Fermi-GBM ($<$ 1 deg uncertainties) and Fermi-LAT GRBs making sure to remove all duplicate detections. We also combine all three catalogs together. The most significant results are shown in Tab.~\ref{tab:FRB-GRB-SEP}. We find several matches with small separations and small chance probabilities $P_{1}$. One of them is the pair FRB\,20171209A - GRB\,110715A found in~\citep{Wang_2020}. The fact that we used a bigger sample and a slightly different approach yields to slightly different chance probabilities. 



\section{Discussion and conclusions}
\label{sec:Discussion}

\begin{figure*}[!th]
  \centering
  \begin{minipage}[b]{0.40\textwidth}
    \includegraphics[width=\textwidth]{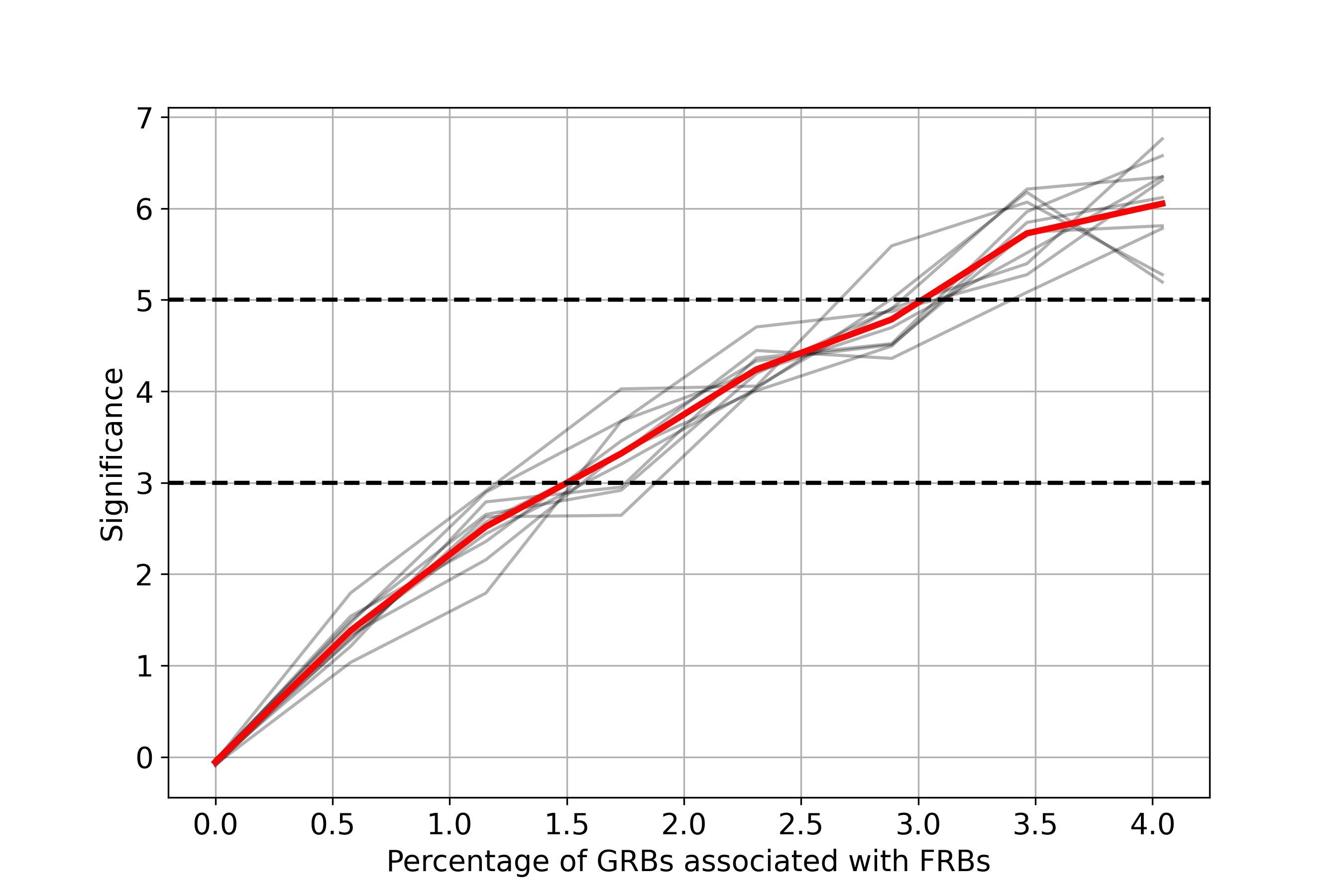}
  \end{minipage}
    \begin{minipage}[b]{0.40\textwidth}
    \includegraphics[width=\textwidth]{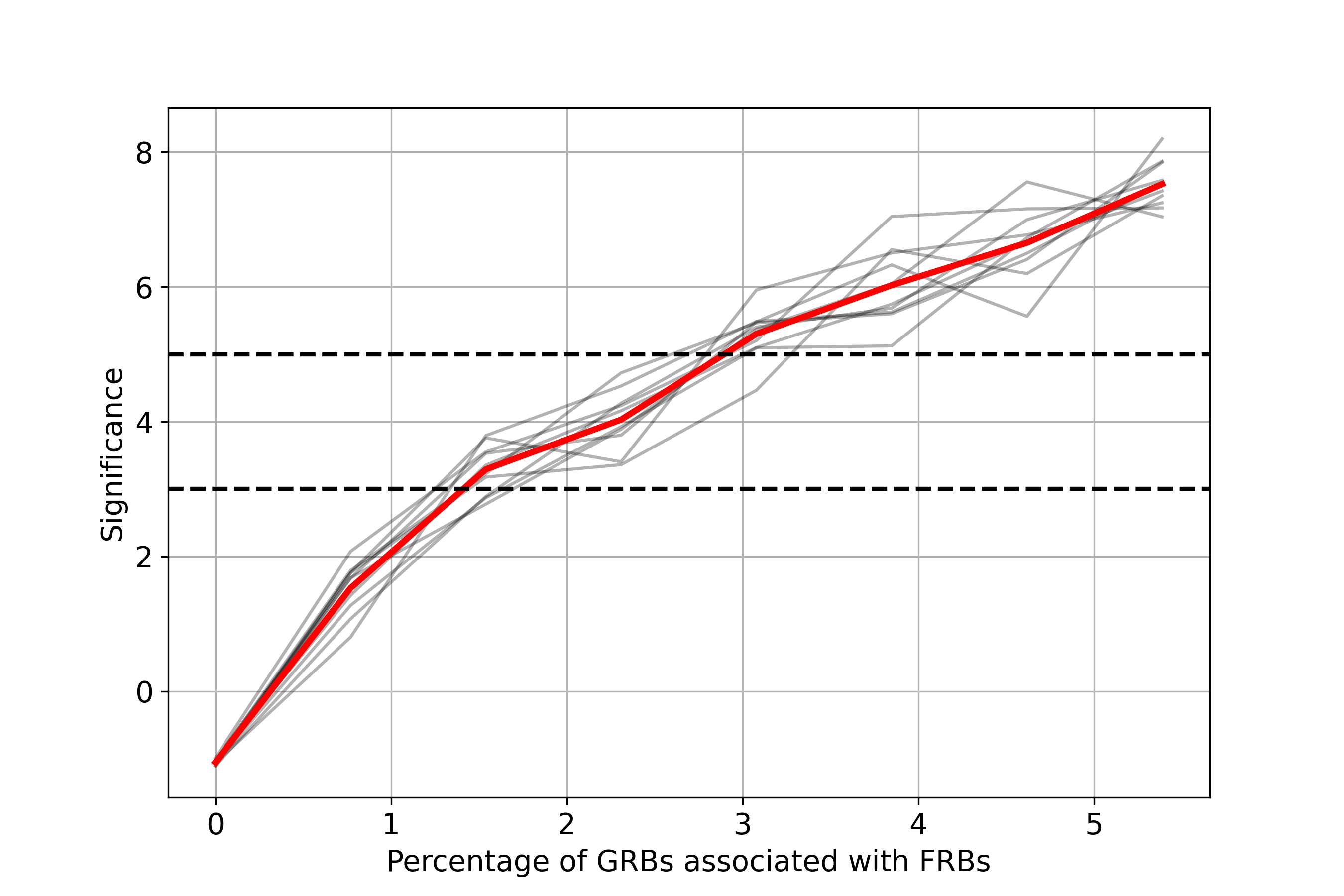}
  \end{minipage}
     \begin{minipage}[b]{0.40\textwidth}
    \includegraphics[width=\textwidth]{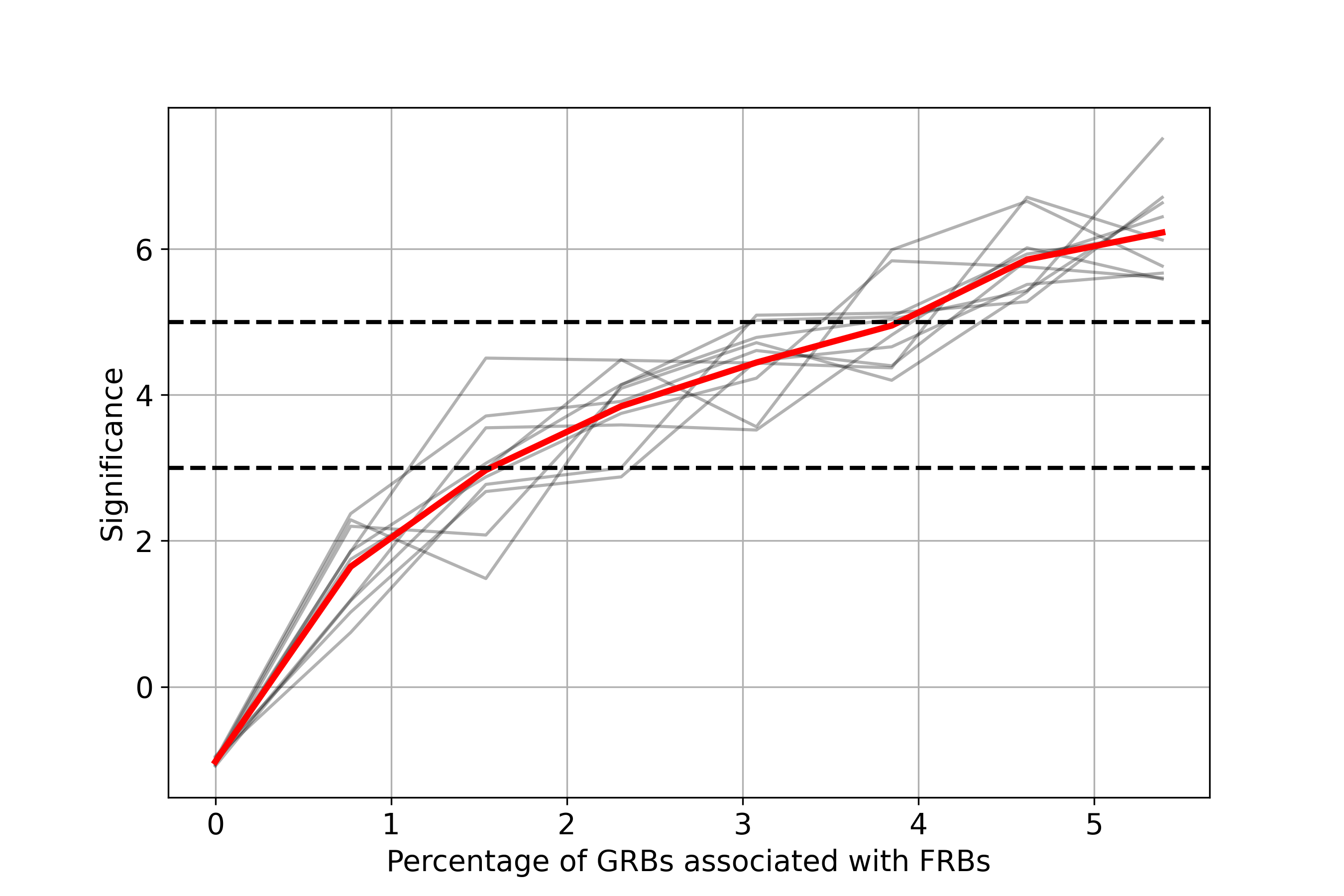}
    \end{minipage}
    \begin{minipage}[b]{0.40\textwidth}
    \includegraphics[width=\textwidth]{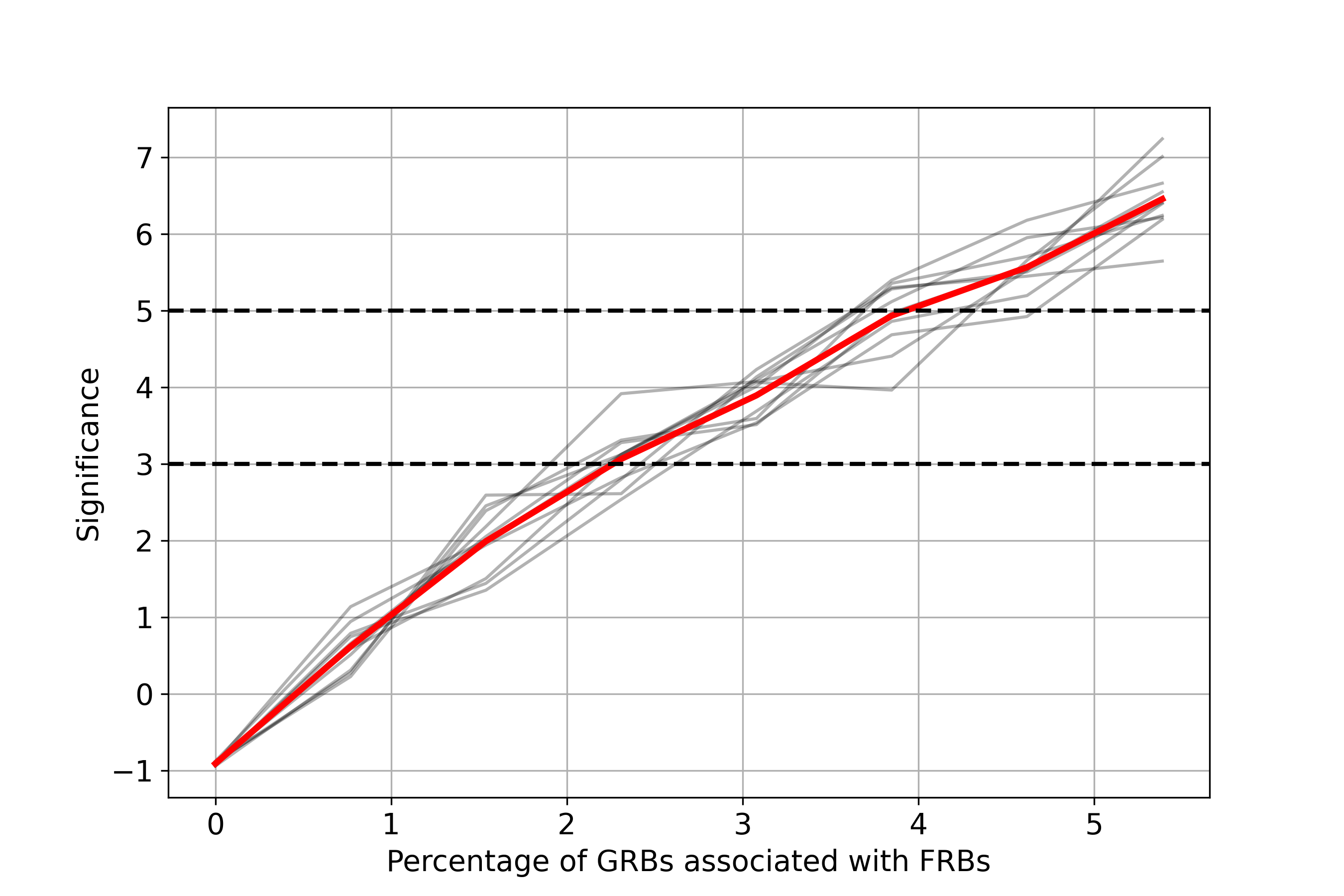}
      \end{minipage}
     \caption{Sensitivity curves showing the significance vs the percentage of injected fake GRBs with FRB positions for the Swift-BAT (upper left), Fermi-GBM ($<$1 deg uncertainty, upper right), Fermi-LAT (lower left) and the combination of all three GRB catalogues (lower right). The black curves show the result of individual sensitivity studies repeated 10 times for each case. The red curves are the average of the black curves for each case.}
     \label{fig:sens_frb_grb}
\end{figure*}

For the population study, where we considered FRB and GRB population as a whole, the histograms show that the signal in the first few bins does not exceed the background by a significant amount that allows us to claim an association between the two populations. In order to asses the sensitivity of the study in Sec.~\ref{sec:PopulationFRB}, we inject fake GRBs that are spatially associated with FRBs. For that, we consider the position of a random FRB with its localization uncertainty. We generate a new position based on this information and assign it to a random GRB in the GRB catalog. The new GRB position is also generated taking into consideration the localization uncertainties of the GRB instruments. The number of fake GRB injections is increased progressively and the analysis in Sec.~\ref{sec:PopulationFRB} is repeated each time. The entire study is then repeated 10 times. Fig.\ref{fig:sens_frb_grb} shows the sensitivity curves for each of the GRB catalogs, apart and combined. The x-axis shows the percentage of GRBs needed to be associated with FRBs to claim as a significant population association and the y-axis shows the confidence level. From these plots, we conclude that less than 3\%, 5\%, 48\% and 3\% of the FRBs are associated with Swift-BAT, Fermi-GBM, Fermi-LAT, and all GRBs combined respectively with a 95\% confidence level. It is also worth mentioning that we applied this study to the Fermi-GBM GRBs that have localization uncertainties larger than 1 degree. The average uncertainty for these GRBs are larger than 6 degrees. We also applied this study to the Fermi-GBM trigger catalog that consists of gamma-ray transients that are not classified as GRBs. Both studies did not yield any significant detection in their current form. Large uncertainties in these catalogs significantly decreases the sensitivity of the study and prevent us from claiming any detection. This discourages us to investigate further catalogs with large uncertainties such as the BATSE GRB catalogue~\citep{Paciesas_1999} with the current methods. 

In the case-by-case search in Sec.~\ref{sec:CaseByCaseFRB}, several matches between FRBs and GRBs with small chance probabilities were found. However, taking into account the number of trials increases these chance probabilities to 1 for nearly all matches. This concludes that no significant association can be claimed for any of the FRB - GRB pairs. The sensitivity of the radio telescopes like CHIME to increasing declination is a caveat that should be taken into consideration. The localization precision of CHIME seems to deteriorate with increasing altitudes. However, the inclusion of this caveat is beyond the scope of this study.  

Finally, considering a scenario where the GRB is generated in the initial cataclysmic event that created the magnetar and the FRB later on by the interactions of the magnetar and the surrounding nebula, it is possible that the delay between the GRB and the FRB might be longer than a few years. In the case of the repeater FRB\,20121104B, the system is believed to be between 20 and 50 years young, while observations of FRB\,20180916A suggest a system between 200 and 500 years old. The contemporaneity of the GRB and the FRB data used here might be an explanation for the lack of significant association. In that case, the use of older GRB data might be beneficial to test this scenario.


\bibliographystyle{JHEP}
\bibliography{main}


%
%
%

\end{document}